\documentclass[12pt]{article}
\usepackage{amsmath,amssymb,amsfonts,color,graphicx,cite,color,psfrag}

\input paperdef

\graphicspath{{figs/}}

\evensidemargin \oddsidemargin
\allowdisplaybreaks

\begin{document}
\thispagestyle{empty}

\def\thefootnote{\fnsymbol{footnote}}

\begin{flushright}
DCPT/08/38\\ 
IPPP/08/19 
\end{flushright}

\vspace{1cm}

\begin{center}
{\large\sc {\bf 
Charged Higgs Bosons in the MSSM at CMS:\\[.5em]
Discovery Reach and Parameter Dependence}}

\vspace{1cm}

\vspace{0.5cm}

{\sc M.~Hashemi$^{\,1}$%
\footnote{previously at IPM, Tehran, Iran}%
\footnote{
email: Majid.Hashemi@cern.ch
}%
, S.~Heinemeyer$^{\,2}$%
\footnote{
email: Sven.Heinemeyer@cern.ch
}%
, R.~Kinnunen$^{\,3}$%
\footnote{
email: Ritva.Kinnunen@cern.ch
}%
, \\[.5em] 
A.~Nikitenko$^{\,4}$%
\footnote{
email: Alexandre.Nikitenko@cern.ch
}%
~and G.~Weiglein$^{\,5}$%
\footnote{
email: Georg.Weiglein@durham.ac.uk
}%
}

\vspace*{0.5cm}

$^1$ Universiteit Antwerpen, G.U.238, Groenenborgerlaan 171,
2020 Antwerpen, Belgium

\vspace*{0.1cm}

$^2$ Instituto de Fisica de Cantabria (CSIC-UC), Santander, Spain

\vspace*{0.1cm}

$^3$ Helsinki Institute of Physics, Helsinki, Finland

\vspace*{0.1cm}

$^4$ Imperial College, London, UK; on leave from ITEP, Moscow, Russia

\vspace*{0.1cm}

$^5$ IPPP, University of Durham, Durham DH1~3LE, UK

\end{center}

\vspace{1cm}

\begin{abstract}
The search for MSSM Higgs bosons will be an important goal at the
LHC. In order to analyze the search reach of the CMS experiment for the
charged MSSM Higgs bosons, we combine the latest results for the
CMS experimental sensitivities based on full simulation studies with 
state-of-the-art theoretical predictions of MSSM Higgs-boson production
and decay properties.
The experimental analyses are done assuming an integrated luminosity of 
30~\ifb\ for the two cases $\MHp < \mt$ and $\MHp > \mt$.
The results are interpreted as 5$\,\si$ discovery contours 
in $\MHp$--$\tb$ planes of the MSSM 
for various benchmark scenarios. We study the
dependence of the $5\,\si$ contours on the variation of the relevant
SUSY parameters. Particular emphasis is put on analyzing the
variation of the discovery contours with the Higgs mixing parameter~$\mu$.
The variation of $\mu$ can shift the prospective discovery reach 
in $\tb$ by up to $\De\tb~=~40$.
\end{abstract}

\def\thefootnote{\arabic{footnote}}
\setcounter{footnote}{0}
\setcounter{page}{0}

\newpage


\section{Introduction}

One of the main goals of the LHC is the identification of the mechanism
of electroweak symmetry breaking. The most frequently investigated
models are the Higgs mechanism within the Standard 
Model (SM) and within the Minimal Supersymmetric Standard Model
(MSSM)~\cite{mssm}. Contrary to the case of the SM, in the MSSM 
two Higgs doublets are required.
This results in five physical Higgs bosons instead of the single Higgs
boson in the SM. These are the light and heavy $\cp$-even Higgs bosons, $h$
and $H$, the $\cp$-odd Higgs boson, $A$, and the charged Higgs bosons,
$H^\pm$.
The Higgs sector of the MSSM can be specified at lowest
order in terms of the gauge couplings, the ratio of the two Higgs vacuum
expectation values, $\tb \equiv v_2/v_1$, and the mass of the $\cp$-odd
Higgs boson, $\MA$ (or $\MHp$, the mass of the charged Higgs boson).
Consequently, the masses of the $\cp$-even neutral and the charged Higgs
bosons are dependent quantities that can be
predicted in terms of the Higgs-sector parameters, e.g.\
$\MHp^2 = \MA^2 + \MW^2$, where $\MW$ denotes the mass of the $W$~boson.
The same applies to
the production and decay properties of the MSSM Higgs bosons%
\footnote{If the production or decay involves SUSY particles at
  tree-level, also other MSSM parameters enter the prediction at lowest
  order.}%
.~Higgs-phenomenology
in the MSSM is strongly affected by higher-order corrections, in
particular from the sector of the third generation quarks and squarks,
so that the dependencies on various other MSSM parameters can be
important, see e.g.\ \citeres{PomssmRep,habilSH,mhiggsAWB} for reviews.

Searches for the charged Higgs bosons of the MSSM (or a more general
Two Higgs Doublet Model (THDM)) have been carried out at
LEP~\cite{LEPchargedHiggsPrel}, yielding a bound of
$\MHp \gsim 80 \gev$~\cite{LEPchargedHiggsProc,LEPchargedHiggs}.
The Tevatron placed additional bounds on the MSSM parameter space from
charged Higgs-boson searches, in particular at large $\tb$ and low
$\MA$~\cite{Tevcharged}. At the LHC the charged Higgs bosons will be
accessible best at large $\tb$ up to $\MA \lsim 800 \gev$
\cite{atlastdr,cmstdr,benchmark3}. At the ILC, for 
$\MHp \lsim \sqrt{s}/2$ a high-precision determination of the charged
Higgs boson properties will be
possible~\cite{tesla,orangebook,acfarep,Snowmass05Higgs}.

The prospective sensitivities at
the LHC are usually displayed in terms of the parameters $\MA$ and $\tb$
(or $\MHp$ and $\tb$) that characterize the MSSM Higgs sector at lowest
order. The other MSSM 
parameters are conventionally fixed according to certain benchmark
scenarios~\cite{benchmark2}. 
The respective LHC analyses of the $5\,\si$ discovery contours for the
charged Higgs boson are given in \citere{HchargedATLAS} for
ATLAS and in \citeres{lightHexp,heavyHexp} for CMS. 
However, within these analyses the variation with relevant SUSY
parameters as well as possibly relevant loop corrections in the Higgs
production and decay~\cite{benchmark3} have been neglected. 

We focus in this paper on the $5\,\si$ discovery contours for the
charged MSSM Higgs boson 
for the two cases $\MHp < \mt$ and $\MHp > \mt$, 
within the $\mhmax$~scenario and the no-mixing
scenario~\cite{benchmark2,benchmark3} (i.e.\ we concentrate on the
$\cp$-conserving case).
They are obtained by using the latest CMS
results~\cite{lightHexp,heavyHexp} derived in a model-independent
approach, i.e.\ making no assumption on the Higgs boson production
mechanism or decays. However, the detection relies on the decay mode
of the charged Higgs bosons to $\tau\nu_\tau$. Furthermore only SM
backgrounds have been assumed. 
These experimental results are combined with up-to-date theoretical
predictions for charged Higgs production and decay in the MSSM, taking
into account also the decay to SUSY particles that can in principle
suppress the branching ratio of the charged Higgs boson decay to
$\tau\nu_\tau$.

For the interpretation of the exclusion bounds and prospective discovery
contours in the benchmark scenarios it is important to assess how
sensitively the results depend on those parameters that have been fixed
according to the benchmark prescriptions. In \citeres{benchmark3,cmsHiggs}
this issue has been analyzed for the neutral heavy MSSM Higgs bosons,
and it has been found that the by far largest effect arises from the
variation of the Higgs-mixing parameter~$\mu$. 
Consequently, we investigate how the 
5$\,\si$ discovery regions in the $\MHp$--$\tb$ plane 
for the charged MSSM Higgs boson obtainable with the CMS experiment at
the LHC are affected by a variation of the 
mixing parameter~$\mu$.


\section{Experimental analysis}
\label{sec:exp}

The main
production channels at the LHC are
\BE
pp \to t\bar t \; + \; X, \quad
t \bar t \to t \; H^- \bar b \mbox{~~or~~} H^+ b \; \bar t~
\label{pp2Hpm}
\EE
and
\BE
gb \to H^- t \mbox{~~or~~} g \bar b \to H^+ \bar t~.
\label{gb2Hpm}
\EE
The decay used in the analysis to detect the charged Higgs boson is
\BE
H^\pm \; \to \; \tau \nu_\tau \; \to \; {\rm hadrons~}\nu_\tau. 
\label{Hbug}
\EE
The analyses described below correspond to 
CMS experimental sensitivities based on full simulation studies, 
assuming an integrated luminosity of 30~\ifb.
In these analyses a top quark mass of $\mt = 175 \gev$ has been
assumed. 


\subsection{The light charged Higgs Boson}
\label{sec:lightHpm}

The ``light charged Higgs boson'' is characterized by $\MHp < \mt$. 
The main production channel is given in \refeq{pp2Hpm}. Close to
threshold also \refeq{gb2Hpm} contributes. The relevant (i.e.\
detectable) decay channel is given by \refeq{Hbug}.
The experimental analysis, based on 30~\ifb\ collected with CMS, is
presented in \citere{lightHexp}. The events were required to be
selected with the single lepton trigger, thus exploiting the
$W \to \ell \nu$ decay mode of a $W$~boson from the decay of 
one of the top quarks in \refeq{pp2Hpm}.

The total number of events leading to final states with the signal
characteristics is evaluated, including their respective experimental
efficiencies. The various channels and the corresponding efficiencies
can be found in \refta{tab:lightHp}. The efficiencies are given for
$\MHp = 160 \gev$, but vary only insignificantly over the parameter
space under investigation. 
The number of signal-like events is evaluated as the sum of 
background and Higgs-boson signal events, 
\begin{align}
N_{\rm ev} =& \;N_{\rm background} 
               \mathrm{(from~the~processes~in~\refta{tab:lightHp})} \non \\
          &+ \cL \times \si(pp \to t \bar t + X) 
                 \times \br(t \to H^\pm b)
                 \times \br(H^\pm \to \tau \nu_\tau) \\
          &\mbox{}\hspace{41.5mm} \times \br(\tau \to \mbox{hadrons})
                 \times \mbox{exp.\ eff.}~, \non 
\end{align}
where $\cL$ denotes the luminosity, and the experimental efficiency
is given in \refta{tab:lightHp}.
A $5\,\si$ discovery can be achieved if a parameter point results in
more than 5260~events (with 30~\ifb).\\
\newpage
\noindent
We furthermore used 
\begin{align}
\br(W^\pm \to \ell \nu_\ell) &~= ~0.217 ~~(\ell = \mu, e), \non \\ 
\br(W^\pm \to \tau \nu_\tau) &~= ~0.1085 , \non \\
\br(W^\pm \to \mbox{jets}) &~= ~0.67 , \\
\br(\tau \to \mbox{hadrons}) &~= ~0.65 . \non
\end{align}
The next-to-leading order LHC cross section  for top quark pairs is
taken to be 840~pb~\cite{sigmatt}. 
For the $W^\pm$+3 jets background the leading 
order cross section for the process $pp \to W^{\pm} + \rm 3~jets$, 
$W^{\pm} \to \ell^{\pm} \nu$ ($\ell=e,~\mu$) of 840~pb was used, 
as given by the MadGraph~\cite{MadGraph} generator.

\begin{table}[htb!]
\renewcommand{\arraystretch}{1.5}
\BC
\begin{tabular}{|c|c|} \hline
channel & exp.\ efficiency \\ \hline\hline
$pp \to t \bar t +X,\; t \bar t \to H^+ b \; \bar t 
                 \to (\tau^+ \bar{\nu}_\tau) \; b \; (W^- \bar b)$;  
~$\tau \to \mbox{hadrons}$, $W \to \ell \nu_\ell$ & 0.0052 \\
\hline
$pp \to t \bar t +X,\; t \bar t  \to W^+ \; W^- \; b \bar b
                 \to (\tau \nu_\tau) \; (\ell \nu_\ell) \; b \bar b$;
~$\tau \to \mbox{hadrons}$ & 0.00217 \\
\hline
$pp \to t \bar t  +X,\; t \bar t \to W^+ \; W^- \; b \bar b
            \to (\ell \nu_\ell) \; (\ell \nu_\ell) \; b \bar b$ & 0.000859 \\
\hline
$pp \to t \bar t  +X,\; t \bar t \to W^+ \; W^- \; b \bar b
            \to (\mbox{jet jet}) \; (\ell \nu_\ell) \; b \bar b$ & 0.000134 \\
\hline
$pp \to W + \rm 3~jets$, $W \to \ell \nu$ & 0.000013 \\
\hline\hline
\end{tabular}
\EC
\vspace{-1em}
\caption{Relevant signal (first line) and background 
  channels for the light charged Higgs boson and their
  respective experimental efficiencies. The charge conjugated processes
  ought to be included. The efficiency for the charged Higgs production
  is given for $\MHp = 160 \gev$, but varies only insignificantly
  over the relevant parameter space. $\ell$ denotes $e$ or $\mu$.
}
\label{tab:lightHp}
\renewcommand{\arraystretch}{1.0}
\end{table}


\subsection{The heavy charged Higgs Boson}
\label{sec:heavyHpm}

The ``heavy charged Higgs boson'' is characterized by $\MHp \gsim \mt$.
Here \refeq{gb2Hpm} gives the largest contribution to the production cross
section, and very close to 
threshold \refeq{pp2Hpm} can contribute somewhat. The relevant decay
channel is again given in \refeq{Hbug}.
The experimental analysis, based on 30~\ifb\ collected with CMS, has been
presented in \citere{heavyHexp}. The fully hadronic final state 
topology was considered, thus events were selected with the single
$\tau$ trigger at Level-1 and the combined $\tau$-$E_{\rm T}^{\rm miss}$ High
Level trigger. 
The backgrounds considered were $t \bar t$, $W^\pm t$, 
$W^\pm + 3~{\rm jets}$ as well as  QCD multi-jet background.
The $t \bar t$ and QCD multi-jet processes were generated with
PYTHIA~\cite{pythia}, $W^\pm t$ was 
generated with the TopRex generator~\cite{toprex} and 
$W^\pm + 3~{\rm jets}$ with MadGraph~\cite{MadGraph}. 
The production cross sections for the $t\bar t$~background processes were
normalized to the NLO cross sections~\cite{sigmatt}.
The total background amounts (after cuts) to 
$1.7 \pm 1$ events, independently of the charged Higgs boson mass.

\noindent
The number of signal events is evaluated as
\BE
N_{\rm ev} = \cL \times \si(pp \to H^\pm + X) 
                 \times \br(H^\pm \to \tau \nu_\tau)
                 \times \br(\tau \to \mbox{hadrons})
                 \times \mbox{exp.\ eff.}~,
\EE
where $\cL$ denotes the luminosity, and the experimental efficiency
is given in \refta{tab:heavyHp} as a function of $\MHp$.
A $5\,\si$ discovery corresponds to a number of signal events larger
than $14.1$.

\begin{table}[htb!]
\renewcommand{\arraystretch}{1.5}
\BC
\begin{tabular}{|c||cccccc|} 
\hline\hline
$\MHp$ [GeV]            & 171.6 & 180.4 & 201.0 & 300.9 & 400.7 & 600.8 \\ 
\hline
exp.\ eff.\  [$10^{-4}$] & 3.5   & 4.0   & 5.0  & 23    & 32    & 42 \\
\hline\hline
\end{tabular}
\EC
\vspace{-1em}
\caption{Experimental efficiencies for the heavy charged Higgs boson
  detection. 
}
\label{tab:heavyHp}
\renewcommand{\arraystretch}{1.0}
\end{table}

The efficiency for the charged Higgs boson production over the
full mass range considered was evaluated with the PYTHIA~\cite{pythia} 
generator processes 401 ($gg \to tbH^{\pm}$) and 402 ($qq \to tbH^{\pm}$) 
implemented as described in ~\citere{tbH}. 


\section{Calculation of cross section and branching ratios}
\label{sec:theo}

While the phenomenology of the production and decay processes of the
charged MSSM Higgs bosons at the LHC is mainly characterized by
the parameters $\MA$ (or $\MHp$) and $\tb$ that govern the Higgs sector
at lowest 
order, other MSSM parameters enter via higher-order contributions (see
e.g.\ \citere{benchmark3} and references therein), 
and also via the kinematics of Higgs-boson decays into
supersymmetric particles. The other MSSM parameters are usually fixed
in terms of benchmark scenarios. The most commonly used scenarios are
the ``$\mhmax$'' and ``no-mixing'' benchmark 
scenarios~\cite{benchmark2,benchmark3}. According to the
definition of \citere{benchmark2} the $\mhmax$ scenario is given by,
\BEA
\mbox{\underline{$\mhmax:$}} &&
\msusy = 1000 \gev, \quad \Xt = 2\, \msusy, \quad \Ab = \At, \non \\
&& \mu = 200 \gev, \quad M_2 = 200 \gev, \quad \mgl = 0.8\,\msusy~.
\label{mhmax}
\EEA
Here $\msusy$ denotes the diagonal soft SUSY-breaking parameters in the
sfermion mass matrices, $\mt\,\Xt \equiv \mt\, (\At - \mu/\tb)$ is the
off-diagonal entry in the scalar top mass matrix. $A_{t(b)}$ denote the
trilinear Higgs-stop (-sbottom) couplings, $\mu$ is the Higgs mixing
parameter, $\mgl$ the gluino mass, and $M_2$ and $M_1$ denote the soft
SUSY-breaking parameters in the chargino/neutralino sector. 
The parameter $M_1$ is fixed via the GUT relation 
$M_1 = (5\sw^2)/(3\cw^2) \, M_2$.
The no-mixing scenario differs from the $\mhmax$ scenario only in the
definition of 
vanishing mixing in the stop sector and a larger value of $\msusy$,
\BEA
\mbox{\underline{no-mixing:}} &&
\msusy = 2000 \gev, \quad \Xt = 0, \quad \Ab = \At, \non \\
&& \mu = 200 \gev, \quad M_2 = 200 \gev, \quad \mgl = 0.8\,\msusy~.
\label{nomix}
\EEA
The value of the top-quark mass in \citere{benchmark2} was chosen
according to the experimental central value at that time. For our
numerical analysis below, we use 
the value, $\mt = 175 \gev$, see \refse{sec:exp}. 
Using the current value of $\mt = 172.6 \gev$~\cite{mt1726}
would lead to a small shift of the discovery contours right at
threshold, but is insignificant for the qualitative results of this
analysis. 

In \citere{benchmark3} it was suggested that in the search for heavy
MSSM Higgs bosons the $\mhmax$ and no-mixing scenarios, which originally
were mainly designed for the search for the light $\cp$-even Higgs boson
$h$, should be extended by several discrete values of $\mu$ (see below),
\BE
\mu = \pm 200, \pm 500, \pm 1000 \gev ~.
\label{eq:variationmu}
\EE
In our analyses here we focus on $\mu = \pm 200, \pm 1000 \gev$. 

\bigskip
For the calculation of cross sections and branching ratios we use a
combination of up-to-date theory evaluations. The 
interaction of the charged Higgs boson with the $t/b$~doublet can be
expressed in terms of an effective Lagrangian~\cite{deltamb2},
\BE
\label{effL}
\cL = \frac{g}{2\MW} \frac{\mbms}{1 + \db} \Bigg[ 
    \wz \, V_{tb} \, \tb \; H^+ \bar{t}_L b_R \Bigg] + {\rm h.c.}
\EE
Here $\mbms$ denotes the running bottom quark mass including SM QCD
corrections. 
The prefactor $1/(1 + \db)$ in \refeq{effL} arises from the
resummation of the leading $\tb$-enhanced corrections to all orders. 
The explicit
form of $\db$ in the limit of heavy SUSY masses and $\tb \gg 1$
reads~\cite{deltamb1}
\BE
\db = \frac{2\als}{3\,\pi} \, \mgl \, \mu \, \tb \,
                    \times \, I(\msbe, \msbz, \mgl) +
      \frac{\alt}{4\,\pi} \, \At \, \mu \, \tb \,
                    \times \, I(\mste, \mstz, |\mu|) ~.
\label{def:dmb}
\EE
Here $\mste$, $\mstz$, $\msbe$, $\msbz$ denote the $\Stop$ and
$\Sbot$~masses. $\als$ is the strong coupling
constant, while $\alt \equiv h_t^2 / (4 \pi)$ is defined via the top
Yukawa coupling. The analytical expression for $I(\ldots)$ can be found
in \citere{benchmark3}.
Large negative values of $(\mu\,\mgl)$ and $(\mu\,\At)$ (it should be
noted that both
benchmark scenarios have positive $\mgl$ and $\At$) can lead to a
strong enhancement of the  
$H^\pm t b$ coupling, while large positive values lead to a strong
suppression. 
Concerning the $\mhmax$ and the no-mixing benchmark scenarios, 
as discussed in \citeres{cmsHiggs,benchmark3} the $\db$ effects are 
much more pronounced in the $\mhmax$ scenario, where the two terms in
\refeq{def:dmb} are of similar size. In the no-mixing scenario the first
term in \refeq{def:dmb} dominates, while the second term is small. A
further suppression is caused by the larger value of $\msusy$ (see
\refeq{nomix}) in comparison with the $\mhmax$
scenario. Consequently, the total effect of $\db$ is smaller in the
no-mixing scenario (see also the discussion in \citere{benchmark3}). 

For the production cross section in \refeq{pp2Hpm} we use the SM cross
section $\si(pp \to t \bar t) = 840~\rm{pb}$~\cite{sigmatt}%
\footnote{
The corresponding SUSY corrections are small~\cite{sigmattSUSY} and have
been neglected. 
}%
~times the $\br(t \to H^\pm\, b)$ including the $\db$ corrections
described above. 
The production cross section in \refeq{gb2Hpm} is evaluated as given in
\citeres{HpmXSa,HpmXSb}. In addition also the $\db$ corrections of
\refeq{effL} are applied. Finally the $\br(H^\pm \to \tau \nu_\tau)$ is
evaluated taking into account all decay channels, among which the most
relevant are $H^\pm \to tb, cs, W^{(*)}h$. Also possible decays to 
SUSY particles are taken into account. For the decay to $tb$ again
the $\db$ corrections are included.
All the numerical evaluations are performed with the program 
{\tt FeynHiggs}~\cite{feynhiggs,mhiggslong,mhiggsAEC,mhcMSSMlong}, see
also \citere{mhcMSSM2L}.


\section{Numerical analysis}
\label{sec:numanal}

The numerical analysis has been performed in the $\mhmax$~and the
no-mixing scenarios~\cite{benchmark2,benchmark3} for 
$\mu = -1000, -200, +200, +1000 \gev$. 
We separately present the results for the light and the heavy charged
Higgs and finally compare with the results in the CMS PTDR, where the
results had been obtained fixing $\mu = +200 \gev$ and neglecting the
$\db$ corrections, as well as neglecting the charged Higgs-boson decays
to SUSY particles.


\subsection{The light charged Higgs boson}

In \reffi{fig:reachlight} we show the 
results for the $5\,\si$ discovery contours for the light 
charged Higgs boson, corresponding to the experimental
analysis in \refse{sec:lightHpm}, where the charged Higgs boson
discovery will be possible in the areas above the curves shown in
\reffi{fig:reachlight}. 
As described above, the experimental analysis was performed for the
CMS detector and 30~\ifb. The top quark mass is set to $\mt = 175 \gev$. 
The thick (thin) lines correspond to positive (negative) $\mu$, and the
solid (dotted) lines have $|\mu| = 1000 (200) \gev$. 
The curves stop at $\tb = 60$, where we stopped the evaluation of
production cross section and branching ratios. For negative $\mu$ very
large values of $\tb$ result in a strong enhancement of the bottom
Yukawa coupling, and for $\db \to -1$ the MSSM enters a non-perturbative
regime, see \refeq{effL}. 

\begin{figure}[htb!]
\begin{center}
\includegraphics[width=0.45\textwidth]{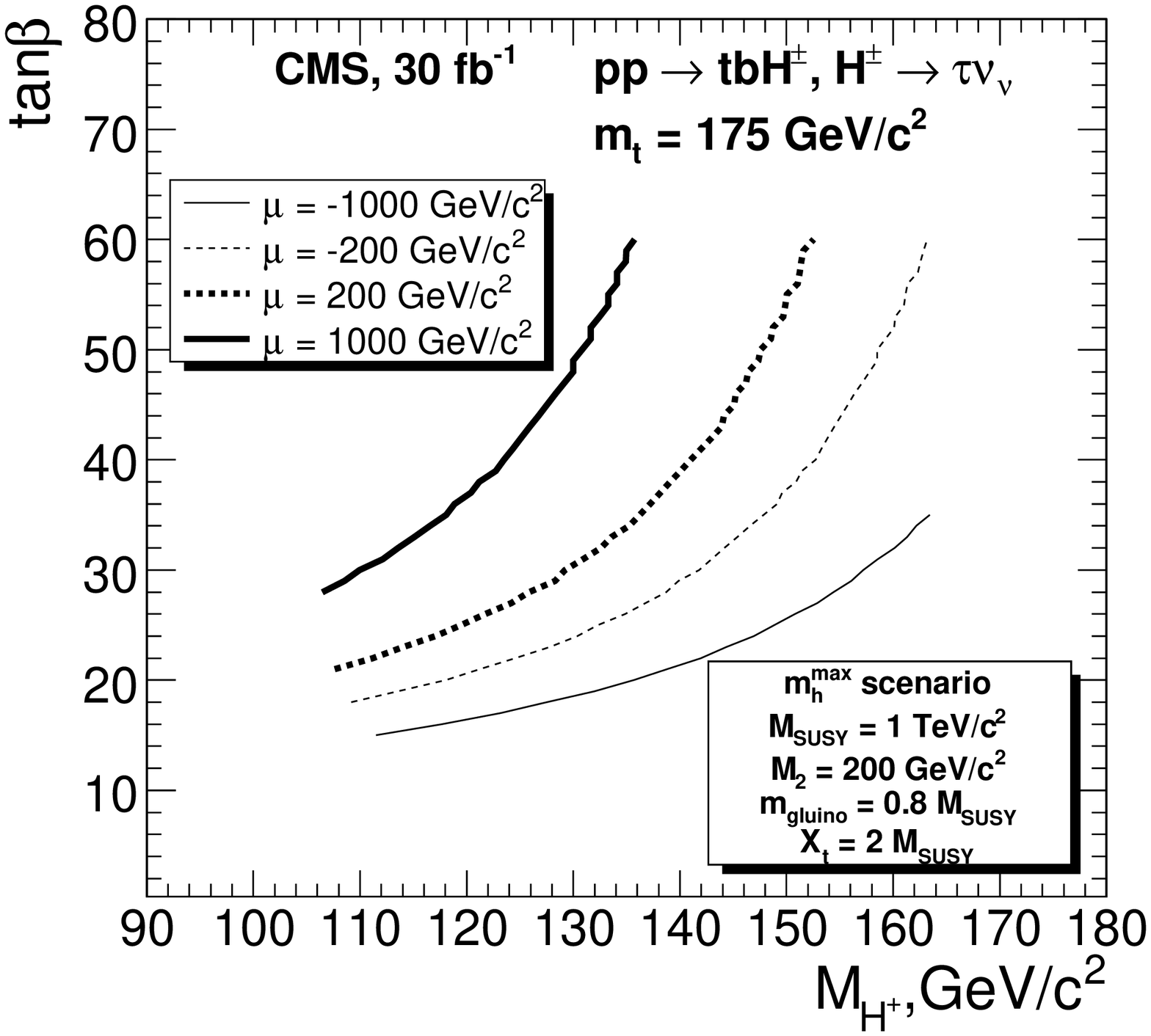}\hspace{1em}
\includegraphics[width=0.45\textwidth]{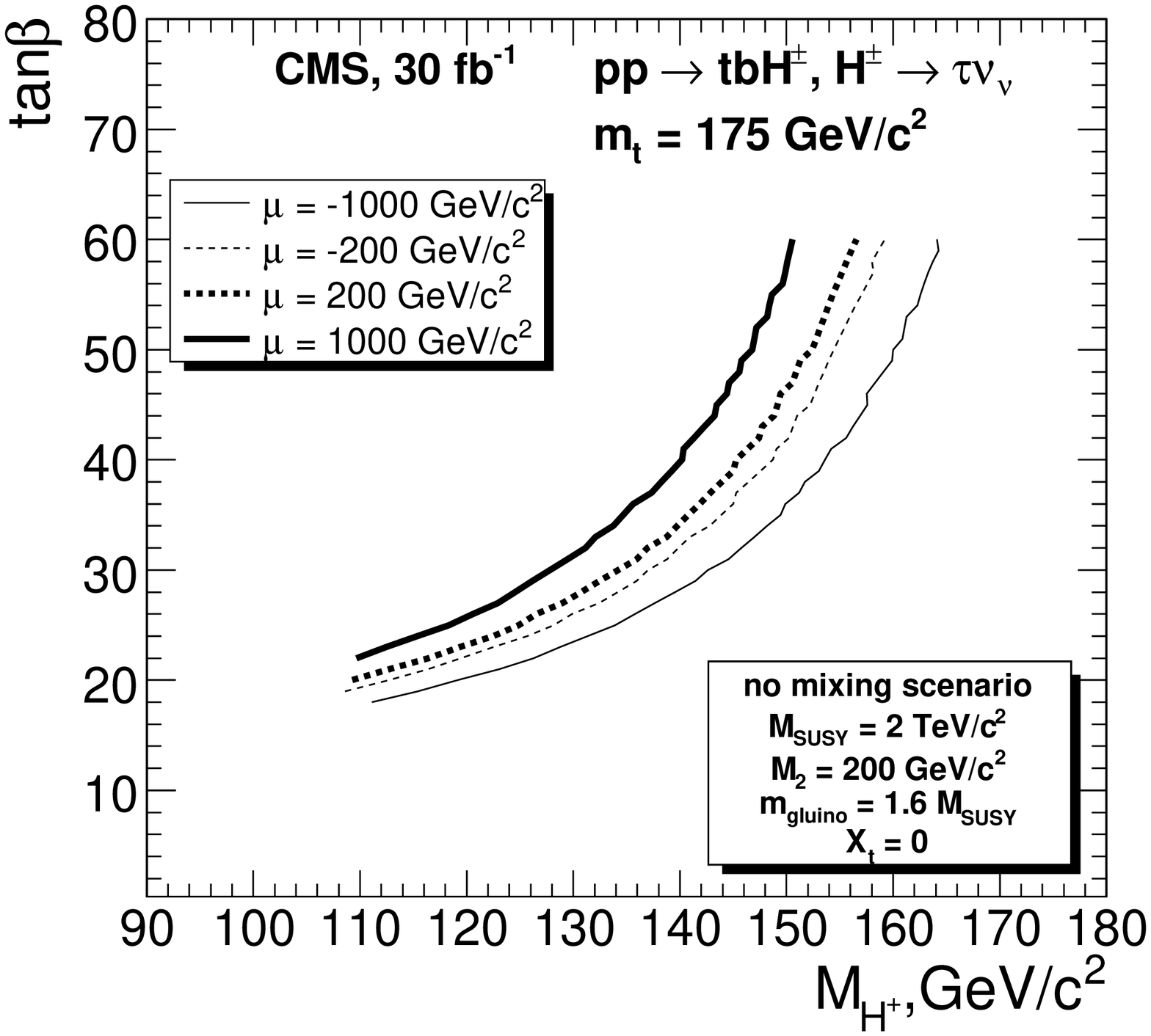}
 \caption{%
Discovery reach for the light charged Higgs boson of CMS with 30~\ifb\ in the 
$\MHp$--$\tb$~plane for the $\mhmax$~scenario (left) and the no-mixing
scenario (right).
}
\label{fig:reachlight}
\end{center}
\end{figure}

Within the $\mhmax$ scenario, shown in the left plot of
\reffi{fig:reachlight}, the search for the light charged Higgs boson covers
the area of large $\tb$ and $\MHp \lsim 130 \ldots 160 \gev$. 
The variation with
$\mu$ induces a strong shift in the $5\,\si$ discovery contours. This
corresponds to a shift in $\tb$ of 
$\De\tb = 15$ for $\MHp \lsim 110 \gev$, rising up to $\De\tb = 40$ for
larger $\MHp$ values. The discovery region is largest (smallest) for 
$\mu = -(+)1000 \gev$, corresponding to the largest (smallest)
production cross section.
The results for the no-mixing scenario are shown in the right plot of
\reffi{fig:reachlight}. The effects of the variation of $\mu$ are much
less pronounced in this scenario, as discussed in \refse{sec:theo}, due
to the smaller 
absolute value of $\db$ (see also the corresponding analysis for neutral
heavy Higgs bosons in \citere{cmsHiggs}). The shift in $\tb$ for
$\MHp = 110 \gev$ is about $\De\tb = 5$ going from $\mu = -1000 \gev$ to
$+1000 \gev$. 
For $\tb = 60$ (where we stop our analysis) the covered $\MHp$ values
range from $150 \gev$ to $164 \gev$.
In this charged Higgs boson mass range for the considered benchmark
scenarios no decay channels into SUSY particles are open, i.e.\ the
observed effects are all due to higher-order corrections, in particular
associated with~$\db$.


\subsection{The heavy charged Higgs boson}

In \reffi{fig:reachheavy} we show the 
results for the $5\,\si$ discovery contours for the heavy
charged Higgs boson, corresponding to the experimental
analysis in \refse{sec:heavyHpm}. The Higgs boson discovery will be
possible in the areas above the curves.%
\footnote{
An analysis in other benchmark scenarios that are in
agreement with the cold dark matter density constraint imposed by WMAP
and other cosmological data~\cite{WMAP} can be found in \citere{ehhow}.}%
~As before, the experimental analysis was performed for the
CMS detector and 30~\ifb. The top quark mass is set to $\mt = 175 \gev$. 
The thick (thin) lines correspond to positive (negative) $\mu$, and the
solid (dotted) lines have $|\mu| = 1000 (200) \gev$. 

\begin{figure}[htb!]
\begin{center}
\includegraphics[width=0.45\textwidth]{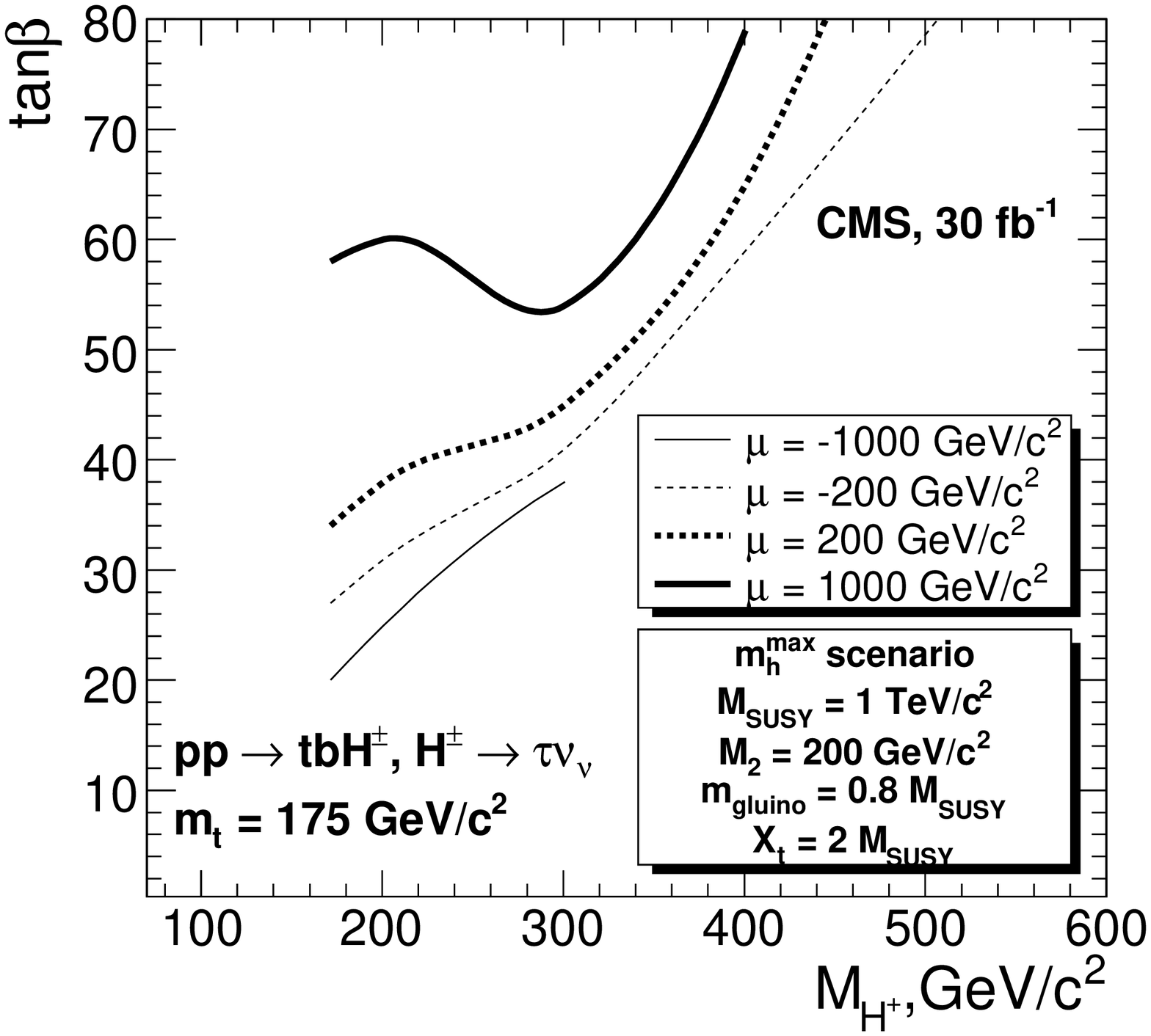}\hspace{1em}
\includegraphics[width=0.45\textwidth]{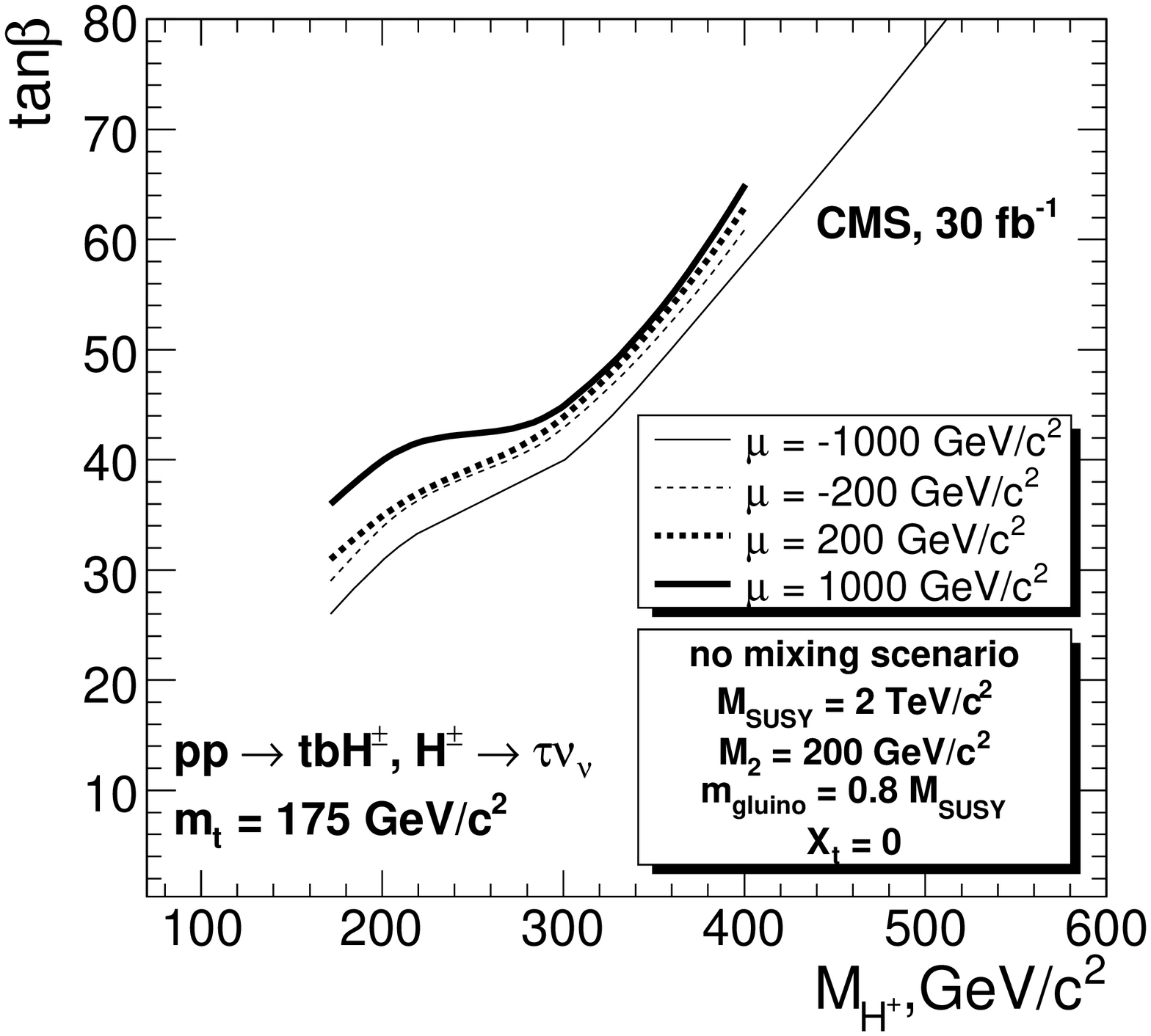}
 \caption{%
Discovery reach for the heavy charged Higgs boson of CMS with 30~\ifb\ in the 
$\MHp$--$\tb$~plane for the $\mhmax$~scenario (left) and the no-mixing
scenario (right).
}
\label{fig:reachheavy}
\end{center}
\end{figure}

The $5\,\si$ discovery regions for the search for heavy charged Higgs
bosons in the $\mhmax$ scenario are shown in the left plot of
\reffi{fig:reachheavy}. For $\MHp = 170 \gev$, where the experimental
analysis stops, we find a strong variation
in the accessible parameter space for $\mu = -(+)1000 \gev$ of $\De\tb = 40$.
It should be noted in this context that close to threshold, where both
production mechanisms, \refeqs{pp2Hpm} and (\ref{gb2Hpm}), contribute,
the theoretical 
uncertainties are somewhat larger than in the other regions. 
For $\MHp = 300 \gev$ the variation in the $5\,\si$ discovery contours
goes from $\tb = 38$ to $\tb = 54$. For $\mu = -1000 \gev$ and larger
$\tb$ values the bottom Yukawa coupling becomes so large 
that a perturbative treatment would no longer be reliable in this
region, and correspondingly we do not continue the respective curve(s).

The shape of the $\mu = +1000 \gev$ curve has a local minimum at 
$\MHp \approx 300 \gev$ that is not (or only very weakly) present in the other
curves, and that is also not visible in the original CMS analysis in
\citere{heavyHexp} (obtained for $\mu = +200 \gev$, but neglecting the
$\db$ effects). The reason for the local minimum can be traced back to
the strongly improved experimental efficiency going from 
$\MHp = 200 \gev$ to $300 \gev$, see \refta{tab:heavyHp}. The better
efficiency at $\MHp = 300 \gev$ corresponds to a lower required cross
section ($\propto \TQb$) and/or a lower $\br(H^\pm \to \tau \nu_\tau)$
to obtain the same number of signal events. 
On the other hand, going from $\MHp = 200 \gev$ to $300 \gev$ this effect 
is in most cases overcompensated by a decrease of the cross
section due to the increase in $\MHp$. The overcompensation results in
an increase in $\tb$ for the higher $\MHp$ value. 
For $\mu = +1000 \gev$, however, $\db$ is very large, 
suppressing strongly the charged Higgs production cross section as well
as the $\br(H^\pm \to tb)$. The overall effect is a somewhat better
reach in $\tb$ for $\MHp = 300 \gev$ than for $\MHp = 200 \gev$. 

In comparison with the analysis of \citere{benchmark3}, based on the
older CMS analysis given in \citere{heavyHexpold}, several differences
can be observed. The feature of the local minimum is absent in
\citere{benchmark3}, the variation of the $5\,\si$ discovery contours
with $\mu$ is weaker, and the effect of the decay of the charged Higgs
boson to a chargino and a neutralino is more pronounced in
\citere{benchmark3}. The reason for these differences is the strongly
reduced discovery region in the new CMS analysis~\cite{heavyHexp}
employed here as compared to the old CMS analysis~\cite{heavyHexpold} 
used in \citere{benchmark3}. The reach in $\tb$ is worse by
$\sim 15 (30)$ for $\MA = 200 (400) \gev$ in the new analysis.%
\footnote{
The old analysis uses $\mu = -200 \gev$~\cite{heavyHexpold}, while the
new analysis set $\mu = +200 \gev$~\cite{heavyHexp}. However, since the
$\db$ corrections are neglected in \citeres{heavyHexpold,heavyHexp}, 
the effect on the discovery regions should be small.
}%
~Thus, at the substantially worse (i.e.\ higher) $\tb$ values employed
here the $\db$ effects are more pronounced, leading to the local minimum
for $\mu = +1000 \gev$ and to a larger absolute variation in $\tb$ with the
size and the sign of $\mu$, see \refse{sec:theo}.
In the high $\tb$ region furthermore the $\db$ effects dominate over the
impact of the decay of the charged Higgs to charginos and neutralinos. 
As an example, for $\mu = +200 \gev$ and $\MHp = 400 \gev$ the old
analysis in \citere{benchmark3} found that the discovery region starts
at $\tb = 32$, where $\br(H^\pm \to \cha{}\neu{}) \approx 15\%$.
Here we find that the discovery region starts at $\tb = 64$, where
$\br(H^\pm \to \cha{}\neu{}) \approx 3\%$.

The no-mixing scenario is shown in the right plot of
\reffi{fig:reachheavy}. The features are the same as in the $\mhmax$
scenario. However, due to the smaller size of $|\db|$, see
\refse{sec:theo}, they are much less pronounced. The variation in $\tb$
stays at or below the level of $\De\tb = 10$ for the whole range of
$\MHp$.


\subsection{Comparison with the CMS PTDR}

In \reffi{fig:reach} we show the 
combined results for the $5\,\si$ discovery contours for the light and
the heavy charged Higgs boson, corresponding to the experimental
analyses in the $\mhmax$ scenario as presented in the two previous
subsections. They are compared with the results presented in the CMS
PTDR~\cite{cmstdr}. Contrary to the previous sections, we now show the
$5\,\si$ discovery contours in the $\MA$--$\tb$ plane. 
The thick (thin) lines correspond to positive (negative) $\mu$, and the
solid (dotted) lines have $|\mu| = 1000 (200) \gev$. The thickened
dotted (red/blue) lines represent the CMS PTDR results, obtained for
$\mu = +200 \gev$ and neglecting the $\db$ effects.

\begin{figure}[htb!]
\begin{center}
\includegraphics[width=0.60\textwidth]{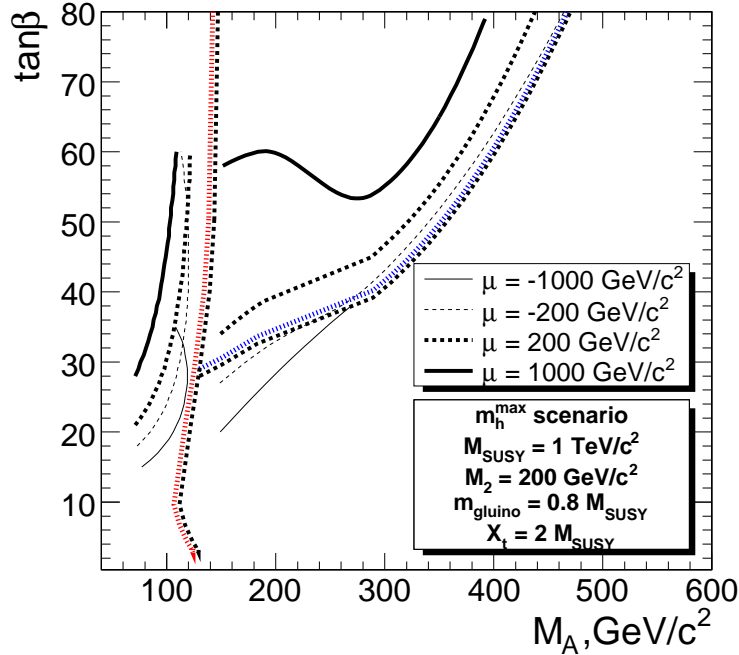}
 \caption{%
Discovery reach for the charged Higgs boson of CMS with 30~\ifb\ in the 
$\MA$--$\tb$~plane for the $\mhmax$~scenario for 
$\mu = \pm 200, \pm 1000 \gev$ in comparison with the results from the CMS
PTDR (thickened dotted (red and blue) lines), obtained for
$\mu = +200 \gev$ and neglecting the $\db$ effects.
}
\label{fig:reach}
\end{center}
\end{figure}

Apart from the variation in the $5\,\si$ discovery contours with the
size and the sign of $|\mu|$, two differences can be observed in the
comparison of the PTDR results to the new results obtained here, i.e.\
including the $\db$ corrections in the production and decay of the
charged Higgs boson as well as taking the decay to SUSY particles into
account. 
For the light charged Higgs analysis the discovery contours are now
shifted to smaller $\MA$ values, for negative $\mu$ even ``bending over''
for larger $\tb$ values. The reason is the more complete inclusion of
higher-order corrections (full one-loop and leading \order{\alt\als}
two-loop) to the relation between $\MA$ and
$\MHp$~\cite{mhcMSSMlong,mhcMSSM2L}.
The second feature is a small gap between the light and the heavy
charged Higgs analyses, while in the PTDR analysis all charged Higgs
masses could be accessed. The gap can be observed best by comparing the
$\mhmax$ scenario in \reffis{fig:reachlight} and \ref{fig:reachheavy}. 
This gap is largest for $\mu = +1000 \gev$ and smallest for 
$\mu = -1000 \gev$, where it amounts only up to $\sim 5 \gev$.
Possibly the heavy charged Higgs analysis strategy exploiting the fully
hadronic final state can be extended to smaller $\MA$ values to
completely close the gap. 
For the interpretation of \reffi{fig:reach} it should be kept in mind
that the accessible area in the heavy Higgs analysis also ``bends over''
to smaller $\MA$ values for larger $\tb$, thus decreasing the visible
gap in \reffi{fig:reach}.


\section{Conclusions}

We have studied the variation of the $5\,\si$ discovery contours for the
search for the charged MSSM Higgs boson with the SUSY parameters.
We combine the latest results for the
CMS experimental sensitivities based on full simulation studies with 
state-of-the-art theoretical predictions of MSSM Higgs-boson properties.
The experimental analyses are done assuming an integrated luminosity of 
30~\ifb\ for the two cases, $\MHp < \mt$ and $\MHp > \mt$.

The numerical analysis has been performed in the $\mhmax$~and the
no-mixing scenarios for $\mu = \pm 200, \pm 1000 \gev$.
The impact of the variation of $\mu$ enters in particular via the
higher-order correction $\db$, affecting
the charged Higgs production cross section and branching ratios. Also
the decays of the charged Higgs boson to SUSY particles have been taken
into account. 
As a general feature, large negative $\mu$ values give the largest
reach, while large positive values yield the smallest $5\,\si$ discovery
areas.

The search for the light charged Higgs boson covers the the area of
large $\tb$ and $\MHp \lsim 160 \gev$. 
The variation with $\mu$ within the $\mhmax$ scenario induces a strong
shift in the $5\,\si$ discovery contours with  
$\De\tb = 15$ for $\MHp = 100 \gev$, rising up to $\De\tb = 40$ for
larger $\MHp$ values. The discovery region is largest (smallest) for 
$\mu = -(+)1000 \gev$, corresponding to the largest (smallest)
production cross section. The effects are similar, but much less
pronounced, in the no-mixing scenario.

The search for the heavy charged Higgs boson reaches up to $\MHp \lsim
400 \gev$ for large $\tb$. 
Within the $\mhmax$ scenario the variation of $\mu$ induces a very
strong shift in the $5\,\si$ discovery contours of up to $\De\tb = 40$
for $\MHp \gsim \mt$. As in the light charged Higgs case, within the
no-mixing scenario the effects show the same qualitative behavior, but
are much less pronounced.

Combining the search for the light and the heavy charge Higgs boson, we
find a small gap, while in the CMS Physics Technical Design Report
analysis all charged Higgs masses could be accessed. 
Possibly the heavy charged Higgs analysis strategy exploiting the fully
hadronic final state can be extended to smaller $\MA$ values to
completely close the gap. This issue deserves further studies.


\subsection*{Acknowledgements}

The work of S.H.\ was partially supported by CICYT (grant FPA~2007--66387).
Work supported in part by the European Community's Marie-Curie Research
Training Network under contract MRTN-CT-2006-035505
`Tools and Precision Calculations for Physics Discoveries at Colliders'.



\end{document}
